\documentclass[a4paper,aps,prb,floatfix,superscriptaddress,notitlepage,twocolumn,nofootinbib]{revtex4-2}

\usepackage{amsmath,amssymb,amsfonts}
\usepackage{graphicx}
\usepackage{bm}
\usepackage{color}
\usepackage[colorlinks=true,linkcolor=blue,citecolor=blue,urlcolor=blue]{hyperref}

\begin{document}

\title{Nonlinear Hall effect as a local probe of plasmonic magnetic hot spots}

\date{\today}

\author{Karina A. Guerrero-Becerra}
\thanks{These authors contributed equally.}
\affiliation{Istituto Italiano di Tecnologia, Graphene Labs, Via Morego 30, I-16163 Genova,~Italy}
\affiliation{Dipartimento di Fisica, Università di Genova, 16146 Genova, Italy}

\author{Andrea Tomadin}
\thanks{These authors contributed equally.}
\affiliation{Dipartimento di Fisica dell'Universit\`a di Pisa, Largo Bruno Pontecorvo 3, I-56127 Pisa, Italy}

\author{Andrea Toma}
\affiliation{Istituto Italiano di Tecnologia, Clean Room Facility, Via Morego 30, I-16163 Genova, Italy}

\author{Remo Proietti Zaccaria}
\affiliation{Istituto Italiano di Tecnologia, Plasmon Nanotechnologies, Via Morego 30, I-16163 Genova, Italy}

\author{Francesco De Angelis}
\affiliation{Istituto Italiano di Tecnologia, Plasmon Nanotechnologies, Via Morego 30, I-16163 Genova, Italy}

\author{Marco Polini}
\affiliation{Dipartimento di Fisica dell'Universit\`a di Pisa, Largo Bruno Pontecorvo 3, I-56127 Pisa, Italy}
\affiliation{\mbox{School of Physics \& Astronomy, University of Manchester, Oxford Road, Manchester M13 9PL, United Kingdom}}
\affiliation{Istituto Italiano di Tecnologia, Graphene Labs, Via Morego 30, I-16163 Genova,~Italy}

\begin{abstract}
Recently developed plasmonic nanostructures are able to generate intense and localized magnetic hot spots in a large spectral range from the terahertz to the visible. 
However, a direct measurement of the magnetic field at the hot spot has not been performed yet, due to the absence of magnetic field detectors that work at those high frequencies and that fit the hot-spot area.
We propose to place a graphene ribbon in the hot spot of a plasmonic nanostructure driven by a laser beam, such that a current is generated due to both the magnetic field at the hot spot and the electric field of the laser.
We demonstrate that a nonlinear Hall voltage, which can be measured by standard electrical means, builds up across the ribbon, making it possible to directly probe the magnetic field at the hot spot.
\end{abstract}

\maketitle

\section{Introduction}

\begin{figure}
(a) \includegraphics[width=0.90\columnwidth]{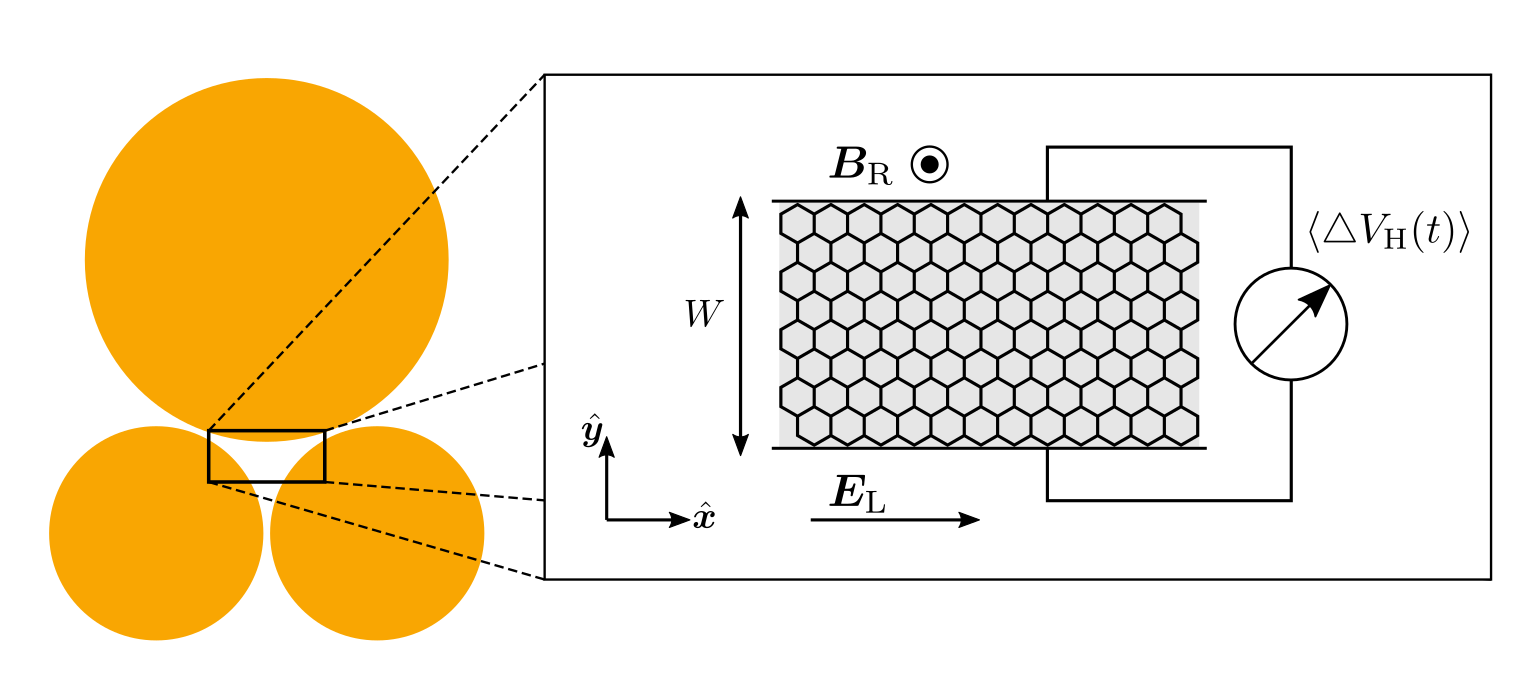}
(b) \includegraphics[width=0.42\columnwidth]{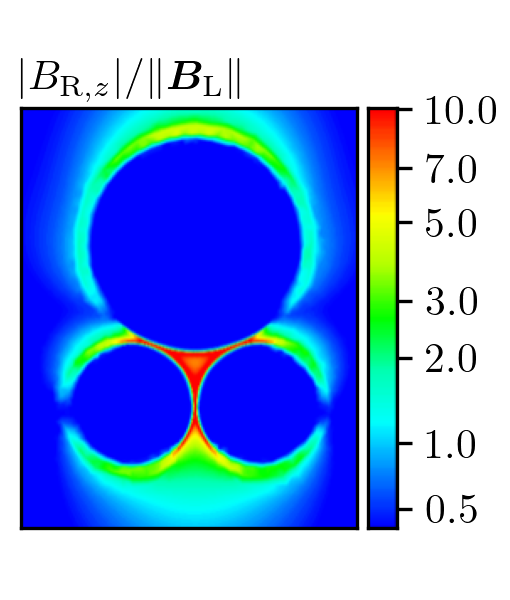}
(c) \includegraphics[width=0.42\columnwidth]{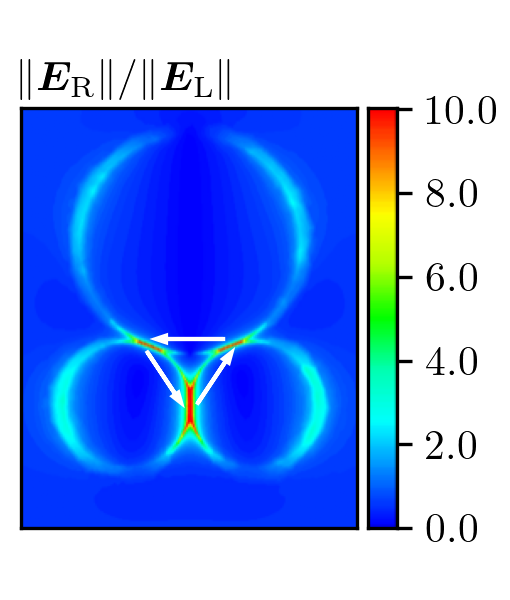}
\caption{(Color online) \label{fig:setup}
(a) Schematics of a graphene ribbon of width $W$ placed in the gap of a nanoassembly composed of three metal disks of different sizes.~\cite{nazir_nanolett_2014}
The Hall voltage $\langle \Delta V_{\rm{H}}(t) \rangle$ between opposite sides of the bar is measured.
(b) The profile of the magnetic field ${\bm B}_{\rm R}$ generated by a plasmonic resonance of the nanostructure.
The magnetic hot-spot is localized in the gap between the three disks.
(c) The profile of the electric field ${\bm E}_{\rm R}$ generated by the plasmonic resonance.
The white arrows schematically represent oscillating dipoles generating the resonant magnetic field.
The electric field is negligible at the position of the magnetic hot-spot.
In both (b) and (c) the fields are shown in units of the corresponding field of the driving laser.
We assume that the electric field ${\bm E}_{\rm L}$ of the driving laser is polarized along $\hat{\bm x}$. }
\end{figure}

Recent advances in nanofabrication have made it possible to build plasmonic nanostructures that are able to generate magnetic field hot spots at optical frequencies spanning from the terahertz (THz) to the visible (VIS) range~\cite{calandrini_nanophotonics_2019} when driven by laser light.
These nanostructures, also known as nanoassemblies, or plasmonic oligomers, are composed of neighboring metallic islands,  with sub-wavelength spatial separation.
Strong near-field coupling between the local plasmonic modes of the islands results in delocalized hybrid plasmonic modes, analog to the delocalized electronic orbitals of aromatic molecules.~\cite{liu_nanolett_2012}
Moreover, small asymmetries in the geometry of the nanoassemblies~\cite{aksyuk_nanoscale_2015} allow coupling between broad electric resonances (which can be excited by a driving laser) and sharp magnetic modes,~\cite{shafiei_naturenano_2013} whose signature appears as a Fano-like resonance in the extinction spectrum.
Appropriate geometries support modes where a circulating displacement current is present in the gaps between the islands, thus reducing Ohmic losses due to conduction currents in the metal.
One such structure, consisting of three metallic disks, has been realized and studied in Ref.~\cite{nazir_nanolett_2014}. (See Fig.~\ref{fig:setup}.)
The structural asymmetry, necessary to couple the radiative electric resonance with the sub-radiant magnetic mode, is obtained by changing the size of one of the disks, relatively to the other two.
In this structure, the magnetic hot spot is confined in the gap between the three disks.
By substituting the larger disk with a moon-shaped island, supporting a quadrupolar-like plasmonic resonance, it has been shown that it is possible to tune the Fano-like resonance frequency without affecting the spatial extent of the hot spot.~\cite{panaro_nanolett_2015}

\vspace{-12pt}

\clearpage

Notwithstanding the progress in the design and fabrication of nanoassemblies able to harvest external radiation into a sub-wavelength near-field, a general and convenient approach to convert the generated AC field into a DC electric signal (i.e.~to ``rectify'' the field) in a spatially-resolved fashion has not been devised yet.
The reasons are twofold.
(i) First, there are no magnetic field gauges that can operate at the required high frequencies and that, at the same time, are small enough to probe the local intensity of the magnetic component of the near-field.
(ii) Second, detectors based on bulk, three-dimensional geometries influence the operation of the nanoassembly with their own conductive or dielectric response.

In this Article we propose to use a graphene ribbon as a detector, exploiting its carriers' \emph{nonlinear} response to rectify the near-field of the nanostructure.
Graphene overcomes \emph{both} difficulties described above.
(i) Coupling between the electromagnetic response of the plasmonic nanostructure and graphene plasmons allows to funnel electromagnetic energy into smaller regions, leading to a spatially-resolved detection. 
(ii) Moreover, the disturbance to the neighboring nanostructure is minimized by graphene's reduced footprint.
Graphene-based devices have been demonstrated to be incredibly versatile photodetectors of radiation in the spectral range where these plasmonic magnetic nanostructures operate, i.e.~from the THz to the VIS.
\cite{bonaccorso_naturephoton_2010,low_acs_2014}
The literature on this topic is vast and we refer the reader to reviews for details.~\cite{koppens_naturenano_2014,tredicucci_ieee_2014}

For definiteness, in this Article we discuss our proposal in the context of the geometry reported in Ref.~\cite{nazir_nanolett_2014}, (see Fig.~\ref{fig:setup},) but we emphasize that our approach is suitable to be adapted to several systems with comparable 2D geometry and spectral features.~\cite{calandrini_nanophotonics_2019}
Fig.~\ref{fig:setup}(a) shows a schematics of the setup.
We consider a graphene ribbon of length $L$ and width $W$ placed at the location of the magnetic hot spot produced by the nanoassembly.
The ribbon is contacted in such a way that the electric potential difference between its edges can be measured.
We assume $L \gg W$ and that the ribbon is uniform in the $\hat{\bm x}$ direction.
The average carrier density $\bar{n}$ in the graphene ribbon is tuned by a metallic back-gate, located at a distance $d$ below the plane where the nanoassembly lies.
The plasmonic response of the disks is driven by a linearly-polarized laser beam impinging orthogonally onto the structure.
Its in-plane electric field is uniform and, we assume, directed along $\hat{\bm x}$: ${\bm E}_{\rm L}(t) = E_{\rm L}(t) \hat{\bm x}$, where $E_{\rm L}(t) = E_{\rm L} \cos(\omega t)$.
(We denote by $\omega = 2 \pi f$ the angular frequency of the field and by $T = 1 / f$ its period.)
The magnetic field ${\bm B}_{\rm R}(t)$ is generated by the resonant plasmonic response of the nanostructure.
At the location of the hot spot, the field ${\bm B}_{\rm R}(t)= B_{\rm R}(t) \hat{\bm z}$ is assumed, for simplicity, to be (i) uniform, (ii) normal to the plane of the structure, and (iii) in phase with the driving electric field, i.e.~$B_{\rm R}(t) = B_{\rm R} \cos{(\omega t)}$.~\cite{footnote_enhancement}
We neglect the magnetic field of the impinging laser beam because it is much smaller than $B_{\rm R}$.
Moreover, we also neglect the electric field ${\bm E}_{\rm R}$ of the Fano-like resonance, responsible for the magnetic response, because it is localized away from the hot spot, where the graphene ribbon is located [cfr.~Fig.~\ref{fig:setup}(b) and (c)].

Here we show that, due to the oscillating magnetic field at the hot spot and the oscillating electric field which drives the plasmonic response of the nanostructure, a finite DC Hall voltage is generated between the edges of the ribbon.
The origin of the Hall voltage resides in the {\it nonlinearity} implicit in the Lorentz force, which mixes the current induced by the electric field with the magnetic field.
Since the two driving fields oscillate at the same frequency, it happens that the Lorentz force is always directed along $\hat{\bm y}$.
Indeed, when the electric field induces a current along $\hat{\bm x}$, the magnetic field is directed along $\hat{\bm z}$; when the induced current is along $-\hat{\bm x}$, the magnetic field is also reversed, pointing towards $-\hat{\bm z}$.
Thus, the graphene ribbon acts as an optical rectifier, yielding a DC signal in response to the local AC magnetic field of the nanostructure, \emph{exploiting} the AC electric field of the driving laser.
From the magnitude of the Hall voltage, the enhancement factor of the nanostructure~\cite{calandrini_nanophotonics_2019} can be determined as a function of the frequency of the driving field. 
The simplicity of our setup makes it a general tool to characterize plasmonic magnetic nanostructures without resorting to indirect methods, such as numerical simulations of extinction spectra.~\cite{nazir_nanolett_2014,panaro_nanolett_2015}

Our Article is organized as follows.
In Sec.~\ref{sec:hallvoltage} we discuss the electric potential which builds up across the graphene ribbon, which we call {\it nonlinear Hall voltage} (NLHV).
In Sec.~\ref{sec:plasmawaves} we show that plasma waves are launched in the graphene ribbon by the joint action of the electric and magnetic field acting on graphene's carriers.
In Sec.~\ref{sec:discussion} we focus on the nonlinear mixing of the electric and magnetic fields, which is responsible for the NLHV, and compare it to related nonlinear transport effects.
In Sec.~\ref{sec:summary} we summarize our main findings.

\section{Theory of the NLHV generated by oscillating driving fields}
\label{sec:hallvoltage}

\subsection{Time-averaged Hall voltage}

In the presence of the driving due to the oscillating electric and magnetic fields, a finite electric potential difference $\Delta V_{\rm H}(t)$ arises between the upper and lower edge of the graphene ribbon.
Our key result is that the DC component, i.e.~$\langle \Delta V_{\rm H}(t) \rangle$ (where $\langle\dots\rangle$ denotes averaging with respect to time $t$) does not vanish, notwithstanding the vanishing time-average of the driving fields.
Indeed, as we show below, the magnitude of such DC signal, the NLHV, is well approximated by the expression
\begin{equation}\label{eq:Vh-2}
\langle \Delta V_{\rm{H}}(t) \rangle  = \frac{W \sigma(\omega) E_{\rm L} B_{\rm R} }{2 \bar{n} e}~,
\end{equation}
where $\sigma(\omega)$ is the frequency-dependent conductivity of graphene's carriers and $-e$ is the electron charge.
The expression for the NLHV contains the product of the electric and magnetic fields, and it is thus nonlinear in the strength of the external driving.
The graphene ribbon works thus as a rectifier for the driving fields, whose periodic oscillations give rise to a constant time-averaged signal $\langle \Delta V_{\rm{H}}(t) \rangle$.
By measuring the NLHV, it is thus possible to obtain the enhancement factor of the magnetic field at the location of the hot spot, at every frequency $f$.

Let us now demonstrate Eq.~(\ref{eq:Vh-2}).
The total electric field in the graphene ribbon is
\begin{equation} \label{eq:etot}
{\bm E}(y,t) \equiv E_{\rm L}(t) \hat{\bm x}+E_{\rm H}(y,t) \hat{\bm y}- \frac{{\bm J}(y,t) \times {\bm B}_{\rm R}(t)}{\bar{n} e}~,
\end{equation}
where, on the right-hand side, we have the external driving field, the Hall field $E_{\rm H}(y, t)$, and the field due to the Lorentz force acting on the electric current density ${\bm J}(y, t)$.
(We use the SI system of units.)
In turn, the electric current density is proportional to the total electric field according to Ohm's law
\begin{equation} \label{eq:const}
{\bm J}(y,t)= \sigma_{0} {\bm E}(y,t)~, 
\end{equation} 
where the constant $\sigma_{0}$ is the electrical conductivity.~\cite{ashcroft_mermin}

The time average of the $y$ component of the electric current density vanishes, i.e.
\begin{equation}\label{eq:jy-av}
\langle J_{y} (y, t)\rangle=0~,
\end{equation}  
where the time-average over a period of the driving is defined by
\begin{equation}\label{eq:average}
\langle g(t) \rangle \equiv \frac{1}{p T} \int_{t_{0}}^{t_{0} + p T} g(t) dt~,
\end{equation}
with integer $p \gg 1$.
To see this, we first recall the continuity equation
\begin{equation}
-e \partial_{t} n(y, t) + \partial_{y} J_{y}(y, t) = 0~,
\end{equation}
where $n(y, t)$ is the carrier density and $-e$ is the electron charge.
Assuming that the system reaches a steady state where the density oscillates periodically under the driving, the time average of the derivative of the carrier density vanishes, leading to $\partial_y \langle J_{y}(y,t) \rangle = 0$. We conclude that the time-averaged $y$ component of the electric current density is uniform.
Since $J_{y}(y,t)$ vanishes at all times at the edges of the ribbon (i.e.~at $y=0$ and $y=W$), Eq.~(\ref{eq:jy-av}) follows.

The $y$ component of Eq.~(\ref{eq:const}) reads
\begin{equation} \label{eq:jy}
J_{y}(y,t)= \sigma_{0} \left [ E_{\rm H}(y,t) + \frac{ J_{x}(y,t) B_{\rm R}(t)}{\bar{n} e} \right ]~.
\end{equation}
Taking the time-average of Eq.~(\ref{eq:jy}) and using Eq.~(\ref{eq:jy-av}), one obtains
\begin{equation}\label{eq:Eh-0}
\langle E_{\rm H}(y, t) \rangle =  - \frac{ \langle J_{x}(y, t) B_{\rm R}(t) \rangle}{\bar{n}e}~,
\end{equation}
The $x$ component of Eq.~(\ref{eq:const}) reads
\begin{equation} \label{eq:jx}
J_{x}(y,t)= \sigma_{0} \left [ E_{\rm L}(t) - \frac{J_{y}(y,t) B_{\rm R}(t)}{\bar{n}e} \right ]~.
\end{equation}
We neglect the second term on the right-hand side, because the field due to the Lorentz force is much smaller than the external driving.
After performing the latter approximation, we can upgrade the constant conductivity $\sigma_{0}$ (which is not adequate beyond the THz range) to the frequency-dependent conductivity $\sigma(\omega)$.
The substitution is made possible by this approximation, because $E_{\rm L}(t)$ oscillates at the single angular frequency $\omega$.
In contrast, the product $J_{y}(y,t)B_{\rm R}(t)$, that we neglect, would introduce oscillations at all the harmonics of the driving frequency, and the linear relation between the current and the field would involve an integral in time, or equivalently, a convolution between $J_{y}(y,\omega)$ and $B_{\rm R}(\omega)$ in the frequency space.~\cite{giuliani_vignale}
(This approximation and the appearance of the harmonics of the driving frequency are extensively discussed in Sec.~\ref{sec:discussion}.)
For the frequency-dependent conductivity, we use the standard Drude expression $\sigma(\omega)= \sigma_{0} / \lbrack 1+(\omega \tau)^2 \rbrack$,~\cite{ashcroft_mermin} where $\sigma_{0} = (\tau e^{2}\bar{n} )/m_{\rm c}$, $\tau $ is the Drude scattering time, and $m_{\rm c}=\hbar \sqrt{\pi \bar{n}/v_{\rm{F}}}$ is the cyclotron mass, with $v_{\rm F}$ the Fermi velocity of graphene's carriers.~\cite{katsnelson_book}

Substituting $J_{x}(y, t)$ into Eq.~(\ref{eq:Eh-0}), we find
\begin{equation} \label{eq:Eh}
\langle E_{\rm H}(y, t) \rangle \approx - \frac{\sigma(\omega) E_{\rm L} B_{\rm R} }{2 \bar{n} e}~.
\end{equation}
The Hall electric potential $V_{\rm{H}}(y,t)$ is related to the field $E_{\rm H}$ by
\begin{equation}
E_{\rm H}(y,t)= - \partial_y V_{\rm{H}}(y,t)~.
\end{equation} 
Finally, the Hall voltage, i.e. the difference between the time-average of the Hall electric potential at the upper and at the lower edge of the graphene ribbon, reads
\begin{eqnarray}\label{eq:Vh-1}
\langle \Delta V_{\rm{H}}(t) \rangle & \equiv & \langle V_{\rm{H}}(y= W,t)\rangle - \langle V_{\rm{H}}(y= 0,t)\rangle \nonumber \\
& = & -\int_{0}^{W} dy \langle E_{H}(y, t)\rangle~. 
\end{eqnarray}
Using Eq.~(\ref{eq:Eh}) one finds Eq.~(\ref{eq:Vh-2}).

\subsection{Linearized hydrodynamic model}
\label{sec:linearhydro}

To assert the validity of the assumptions and approximations made in the derivation of Eq.~(\ref{eq:Vh-2}), we now resort to a hydrodynamic model of the electron system in the presence of the electric field ${\bm E}_{\rm L}(t)$ and the magnetic field ${\bm B}_{\rm R}(t)$. Hydrodynamic models of the electron flow are routinely employed in the modelization of semiconductors.~\cite{vasileska_goodnick}
In graphene, recent experimental results~\cite{bandurin_science_2016, crossno_science_2016, ghahari_prl_2016, krishnakumar_natphys_2017} have demonstrated that a transport regime dominated by hydrodynamic effects is attained in a wide range of carrier densities and temperatures, motivating extensive theoretical investigations, especially focused on the role of the  shear viscosity of the electron fluid.~\cite{tomadin_prl_2014, torre_prb_2015, levitov_natphys_2016, principi_prb_2016, pellegrino_prb_2016, narozhny_annphys_2017, kiselev_prb_2019} 
(For a recent popular review, see e.g.~Ref.~\cite{polini_physicstoday_2020}.)

\begin{figure}
\includegraphics[width=0.98\columnwidth]{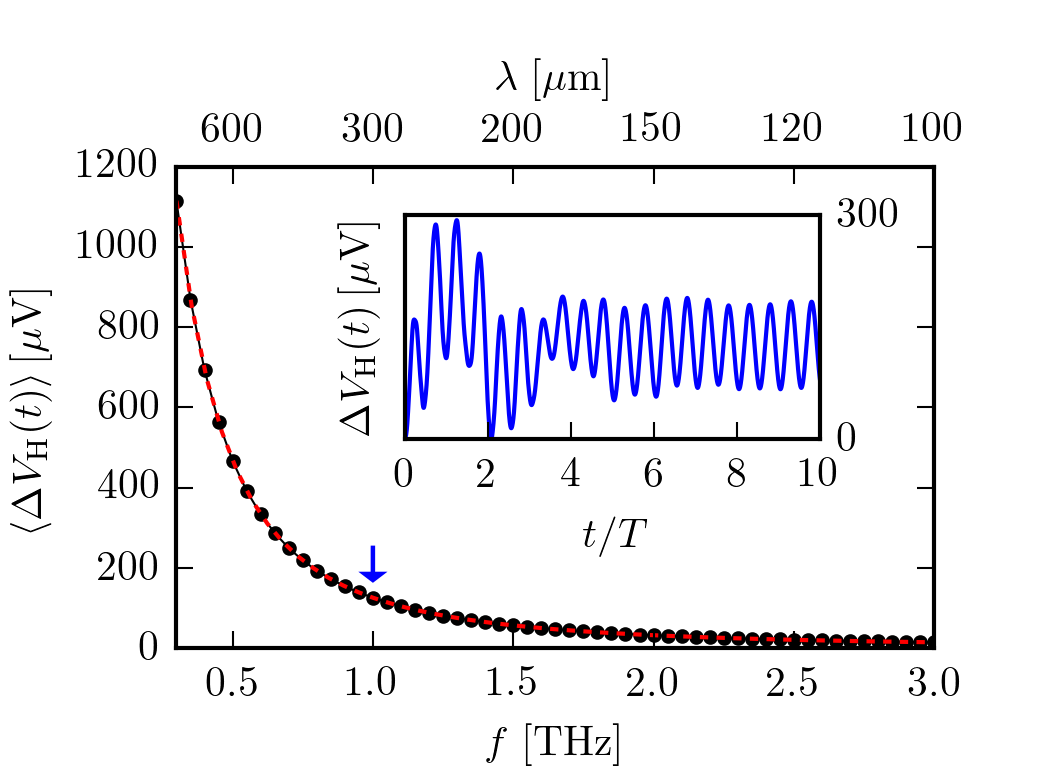}
\caption{\label{fig:hallvoltage}(Color online)
The magnitude of the Hall voltage is shown as a function of the frequency $f$ of the driving laser, within the THz range.
The red dashed line corresponds to the analytical estimate in Eq.~(\ref{eq:Vh-2}) while the black circles to the numerical solution of Eqs.~(\ref{eq:dyn-vis}) for vanishing viscosity $\nu  = 0$.
The inset shows the Hall electric potential as a function of time, at the frequency $f = 1~{\rm THz}$ (indicated by the arrow in the main panel), obtained from the numerical solution, measured in units of the period $T = 1 / f$ of the driving laser.
After an initial transient, the signal oscillates periodically around a non-vanishing average value. }
\end{figure}

\begin{figure}
{(a)}\includegraphics[width=0.98\columnwidth]{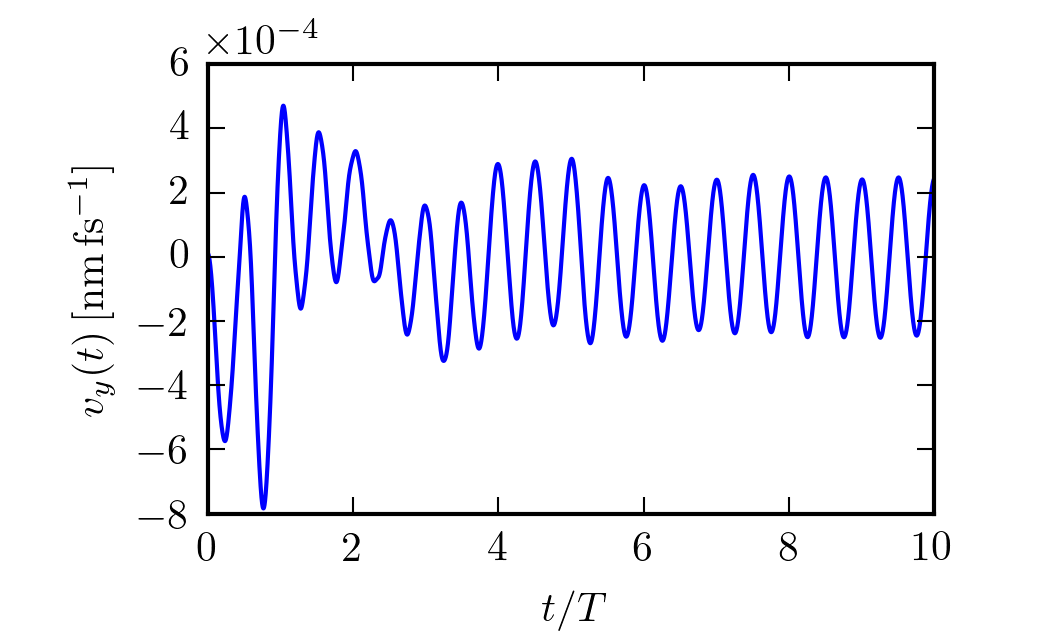}
{(b)}\includegraphics[width=0.98\columnwidth]{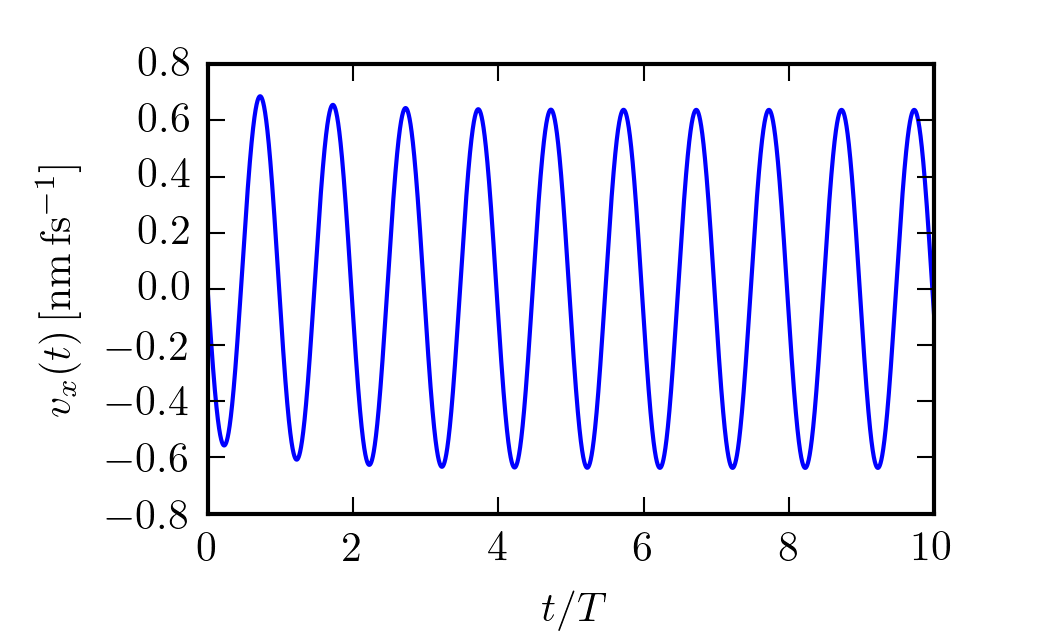}
\caption{\label{fig:velocity}(Color online)
Time-evolution of the components of the velocity ${\bm v}(y, t)$, obtained from the numerical solution of Eqs.~(\ref{eq:dyn-vis}).
(a) After an initial transient, the $y$ component of the velocity oscillates with frequency $2 f$, where $f = 1 / T$ is the  frequency of the external drive.
(b) The $x$ component of the velocity oscillates with the frequency $f$ of the external drive.
The velocity is calculated at position $y = W / 4$. }
\end{figure}

The linearized hydrodynamic model~\cite{footnote_nonlinear} comprises the continuity equation~\cite{landau_book_1987,batchelor_1967}
\begin{subequations}\label{eq:dyn-vis}
\begin{equation}
\partial_t [\delta n(y,t)] + \bar{n} \partial_y  v_y(y,t)=0~,
\end{equation}
and the Navier-Stokes equations~\cite{landau_book_1987,batchelor_1967}
\begin{eqnarray}
\partial_t v_x(y,t) & = & -\dfrac{e E_{\rm L}(t)}{m_{\rm c}}-\dfrac{e}{m_c } v_y(y,t) B_{\rm R}(t) \nonumber \\
& & - \dfrac{v_x(y,t)}{\tau}+ \nu \partial^{2}_y v_x~,
\end{eqnarray}
\begin{eqnarray}
\partial_t v_y(y,t) & = & -\dfrac{e E_{\rm H}(y, t)}{m_{\rm c}}+\dfrac{e}{m_{\rm c}} v_x(y,t) B_{\rm R}(t) \nonumber \\
& &- \dfrac{v_y(y,t)}{\tau}  + \nu \partial^{2}_y v_y~.
\end{eqnarray}
\end{subequations}
Here, $\delta n(y ,t)$ is the carrier density fluctuation on top of the constant value $\bar{n}$, $v_{x}(y, t)$, and $v_{y}(y, t)$ are the Cartesian components of the \emph{fluid element} velocity.
(More precisely, ${\bm v}({\bm r},t)$ is the average velocity of a patch of locally-thermalized electrons.)~\cite{landau_book_1987,batchelor_1967}
The coefficient $\nu$ represents the kinematic viscosity of graphene's carriers.
In this framework, the Hall electric field $E_{\rm H}(y, t)$ directed along $\hat{\bm y}$ arises 
because of the charge density distribution, $\delta \rho(y, t) = -e \delta n(y, t)$, associated to the density fluctuation.
In the absence of a back gate, one should solve the Poisson equation in three dimensions $(\partial_{y}^{2} + \partial_{z}^{2}) V_{\rm H}(y,z,t) = -\rho(y, t) \delta(z) / \epsilon$ (with $\epsilon = \epsilon_{\rm r} \epsilon_{0}$ the average dielectric constant of the surrounding medium) to find the electric potential $V_{\rm H}(y, t) = V_{\rm H}(y, z=0, t)$ on the graphene ribbon produced by the density fluctuations.
However, if the distance at which the back gate is located, $d$, is much smaller than the typical wavelength of the density fluctuations, it is appropriate to use the so-called local-capacitance approximation~\cite{semicond_book}
\begin{equation}\label{eq:lca}
V_{H}(y, t) = \dfrac{e}{C} \delta n(y,t)~,
\end{equation}
where $C = \epsilon / d$ is the capacitance per unit area of the parallel-plate capacitor composed by the graphene ribbon, the underlying back gate, and the dielectric spacer between them.
The linearized hydrodynamic model is complemented by the following boundary conditions: (i) since no current flows through the top and bottom edges of the ribbon, the orthogonal components of the velocity must vanish, i.e.~$v_{y}(y=W,t) = v_{y}(y=0,t)=0$; (ii) for the tangential component, we choose no-slip boundary conditions, i.e.~$v_{x}(y=0,t) = v_{x}(y=W,t) = 0$ (for a discussion of these conditions in 2D electronic systems see Refs.~\cite{torre_prb_2015,pellegrino_prb_2016,kiselev_prb_2019}). 
The integration over time of Eq.~(\ref{eq:dyn-vis}) also requires a set of initial conditions. At $t = 0$, we choose vanishing density fluctuation and a random distribution of velocities with vanishing spatial average, representing thermal excitations.

One can easily see that Ohm's law, Eq.~(\ref{eq:const}), can be obtained from the linearized Navier-Stokes equations (\ref{eq:dyn-vis}), for vanishing viscosity and by neglecting the kinetic terms $\partial_t v_{x,y}(y,t)$ on the left-hand side.
This term can be safely neglected when it is much smaller than the Ohmic friction term $v_{x,y} / \tau$, i.e.~under the condition $\omega \tau \ll 1$.
It is thus expected (as we verify below) that the estimate~(\ref{eq:Vh-2}) deviates from the results of the linearized hydrodynamic model at sufficiently large frequencies only.

\subsection{NLHV as a function of the driving frequency}
\label{sec:numresults}

We now present our results based on the numerical solution of Eqs.~(\ref{eq:dyn-vis}) and compare them to the analytical estimate in Eq.~(\ref{eq:Vh-2}).
We use the following set of parameters (unless otherwise noted): $W = 5~\mu{\rm m}$, $d = 100~{\rm nm}$, $\bar{n}= 10^{11}~{\rm cm}^{-2}$, $\tau = 1~{\rm ps}$, $E_{\rm L}= 1.5 \times 10^{4}~{\rm V}/{\rm m}$, and $B_{\rm R} = 5~{\rm mT}$.
We investigate driving frequencies in the range $0.3~{\rm THz}$ to $50~{\rm THz}$, corresponding to the THz and VIS spectral ranges.
Plasmonic nanostructures operating in the THz range are larger in size, and thus allow an easier placement of the graphene ribbon.~\cite{calandrini_nanophotonics_2019}
Moreover, in the THz range, our proposed setup is of particular practical interest due to the scarcity of other magnetic field gauges.

Fig.~\ref{fig:hallvoltage} shows that the Hall voltage decreases with increasing frequency.
Inspection of Eq.~(\ref{eq:Vh-2}) shows that this decrease is due to the Lorentzian shape of the dynamical conductivity.
The numerical results and the analytical estimate (\ref{eq:Vh-2}) show excellent agreement in the frequency range $f < 3~{\rm THz}$.
The inset shows the difference between the Hall electric potential at the upper and lower edges of the graphene ribbon, as a function of time.
After an initial transient (which depends on the initial randomization of the velocity variables), the signal oscillates at ${\it twice}$ the frequency $f$ of the driving electric field.
This behavior is easy to understand, since it stems from the fact that this quantity is at least of the second order in the driving fields ${\bm E}_{\rm L}(t)$ and ${\bm B}_{\rm R}(t)$.
The signal oscillates around a non-zero value, which is the Hall voltage shown in the main panel.
To exclude the initial transient from the calculation of the time average, we perform the integration in Eq.~(\ref{eq:average}) with $t_{0} = 4 \, T$ and we use a sufficiently long integration window, with $p = 10^{2} - 10^{3}$, to achieve a high accuracy.
Not surprisingly, a similar time evolution is displayed by the component of the velocity parallel to the Hall field, as shown in Fig.~\ref{fig:velocity}(a).
On the contrary, the component of the velocity parallel to the driving electric field, see Fig.~\ref{fig:velocity}(b), oscillates at the {\it same} frequency of the field itself, and shows a negligible initial transient.
This is due to $v_{x}$ being of the first order in ${\bm E}_{\rm L}(t)$.

\begin{figure}
(a)\includegraphics[width=0.98\columnwidth]{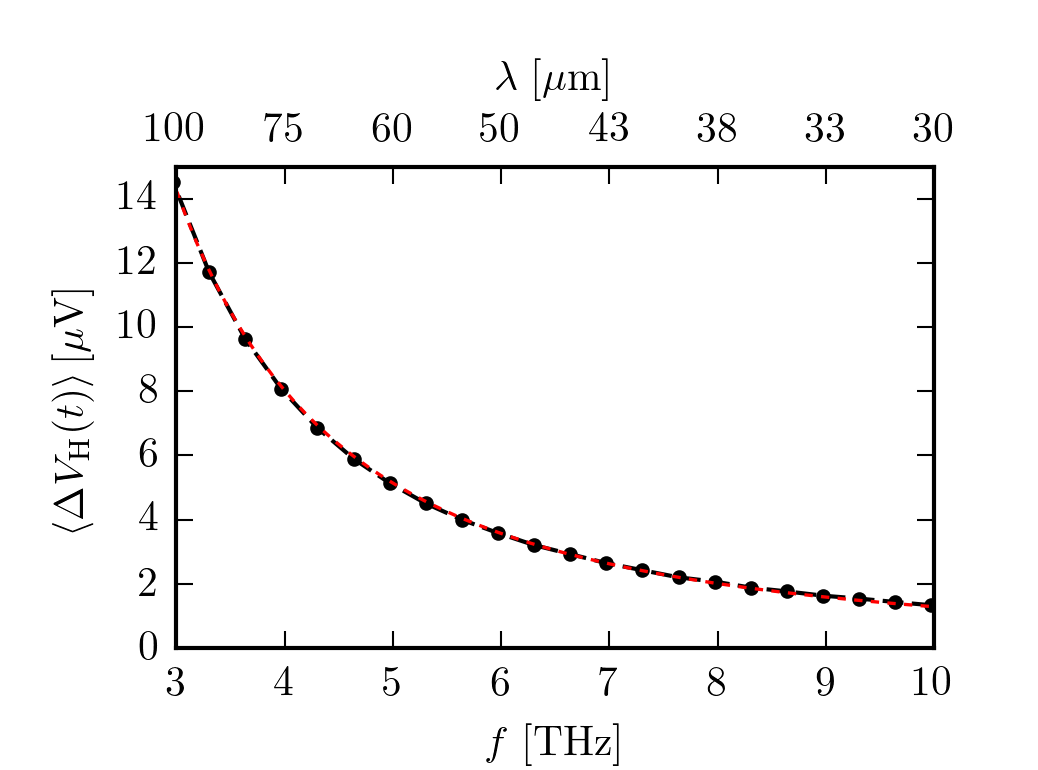}
(b)\includegraphics[width=0.98\columnwidth]{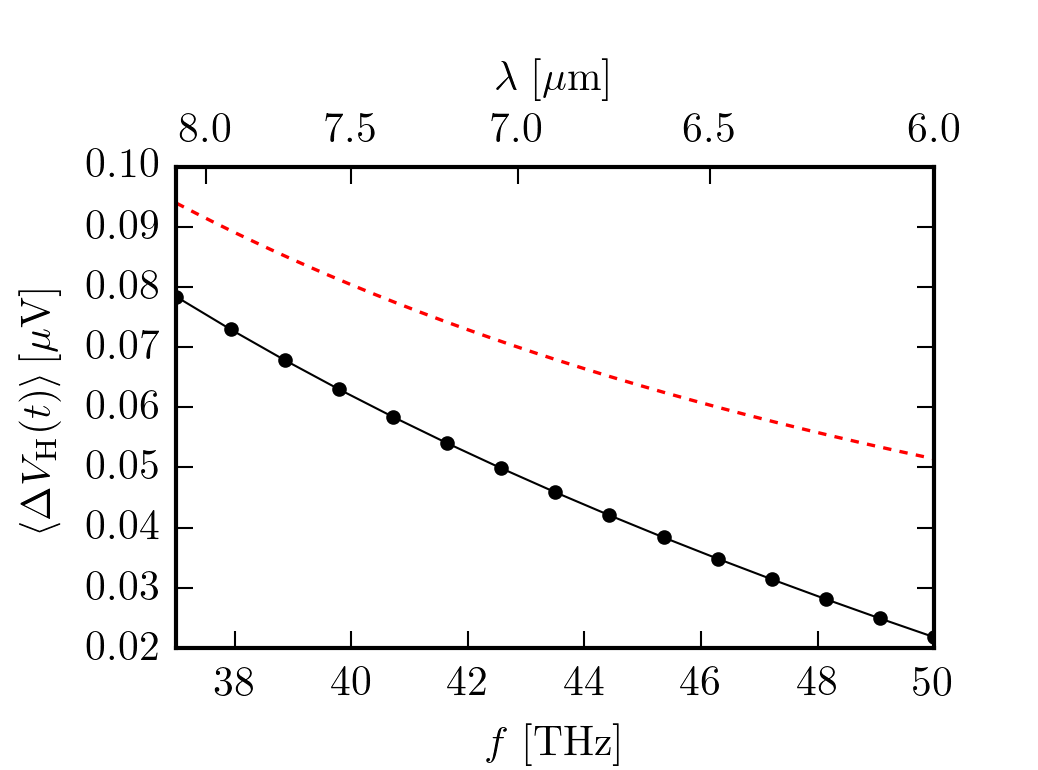}
\caption{\label{fig:largefreqs}(Color online)
Hall voltage at large drive frequencies.
In both panels (a) and (b), the red dashed line corresponds to the analytical estimate in Eq.~(\ref{eq:Vh-2}) while the black circles to the numerical solution of Eqs.~(\ref{eq:dyn-vis}) for vanishing viscosity $\nu  = 0$.
The bottom axis shows the value of the drive frequency $f$ and the top axis the corresponding value of the wavelength $\lambda$.
(a) Frequencies in the far-infrared range.
(b) Frequencies in the MIR range.
Despite its drop in magnitude at large frequencies, the Hall voltage is measurable even in the MIR frequency range. }
\end{figure}

The agreement of the analytical estimate in Eq.~(\ref{eq:Vh-2}) with the numerical results turns out to be excellent in a very wide frequency range, as shown in Fig.~\ref{fig:largefreqs}.
However, deviations up to $50\%$ arise when the frequency is increased to several tens of THz, i.e.~in the mid-infrared (MIR) part of the spectrum.
Since there are metallic nanoassemblies operating at these frequencies,~\cite{calandrini_nanophotonics_2019} the numerical solutions turns out to be important to predict the response of a real device on a quantitative level.

\begin{figure}
\includegraphics[width=\columnwidth]{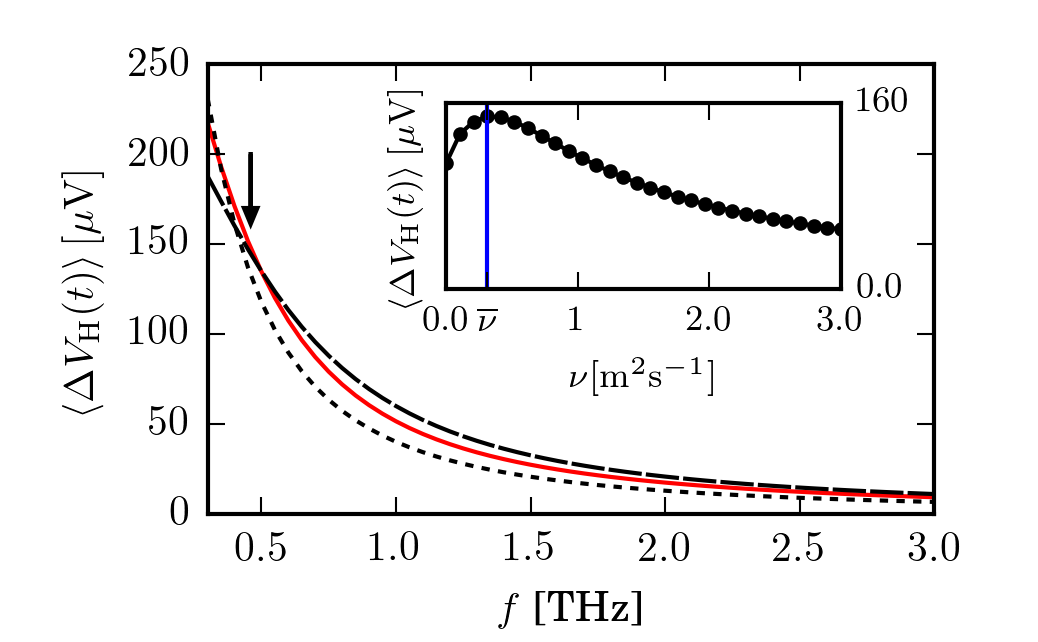}
\caption{\label{fig:viscosity} (Color online)
Hall voltage as a function of the frequency $f$ of the external drive for several values of the viscosity $\nu = 0.1 \, \rm{m^2 s^{-1}}$ (black short-dashed line), $0.3 \, \rm{m^2 s^{-1}}$ (red solid line), and $0.5 \, \rm{m^2 s^{-1}}$ (black long-dashed line), obtained from the numerical solution of Eqs.~(\ref{eq:dyn-vis}).
Differently from the previous figures, here the width of the ribbon is set to $W = 1~ \mu{\rm m}$.
The inset shows the Hall voltage, calculated at frequency $f = 0.45 \, \rm{THz}$ (marked by an arrow in the main panel) as a function of the viscosity $\nu$.
The blue vertical line marks the value $\bar{\nu}$ of the viscosity where the Hall voltage is maximal. }
\end{figure}

Finally, in Fig.~\ref{fig:viscosity}, we show the effect of a finite kinematic viscosity on the Hall voltage.~\cite{principi_prb_2016}
Viscosity is relevant to the dynamics only if the diffusion length ${\cal D}_{\nu} = \sqrt{\nu \tau}$ is comparable to or larger than the width $W$ of the ribbon. For this reason, to use realistic values of the viscosity in graphene~\cite{bandurin_science_2016,krishnakumar_natphys_2017}, Fig.~\ref{fig:viscosity} reports results for a thinner ribbon than the other figures.
Moreover, the hydrodynamic model with finite viscosity is justified if $\omega \tau_{\rm ee} \ll 1$, where $\tau_{\rm ee}$ is the quasiparticle lifetime due to electron-electron collisions.~\cite{torre_prb_2015,principi_prb_2016}
Since in graphene, for $\bar{n}= 10^{11}~{\rm cm^{-2}}$ and at room temperature, $\tau_{ee} \simeq 100~{\rm fs}$,~\cite{torre_prb_2015,principi_prb_2016}, we keep $f < 3~{\rm THz}$ in this figure. We see that the frequency dependence of the Hall voltage is substantially unaltered by a finite viscosity.
This demonstrates that the measurement of the NLHV allows to determine the enhancement factor of the magnetic field at the hot spot, even if the precise value of the kinematic viscosity in the graphene sample is not known.
We notice, however, that the value of the Hall voltage depends on the viscosity in a non-monotonic fashion, as shown in the inset.

\section{Plasma waves in the graphene ribbon}
\label{sec:plasmawaves}

\begin{figure}
{(a)}\includegraphics[width=0.98\columnwidth]{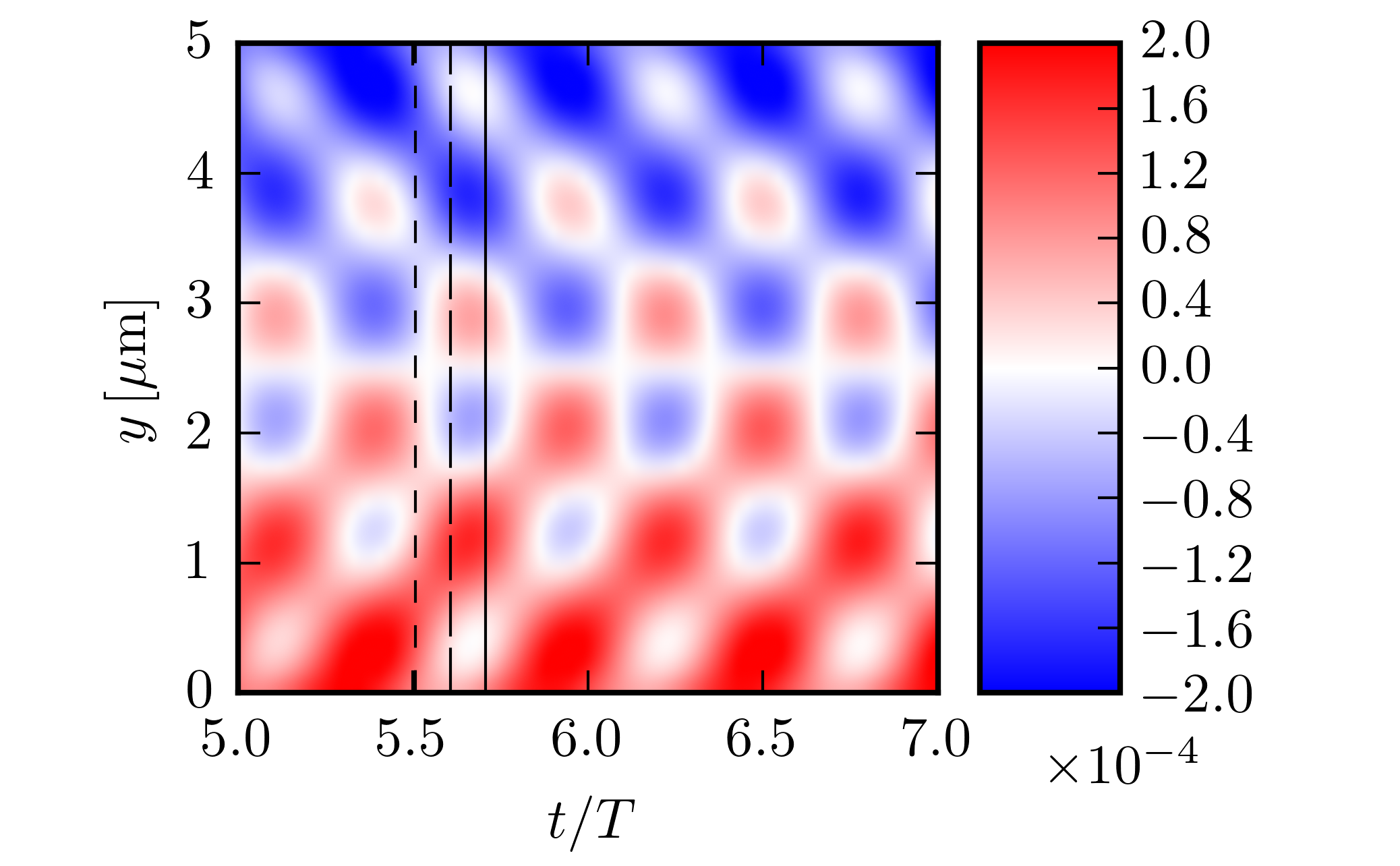}
{(b)}\includegraphics[width=0.98\columnwidth]{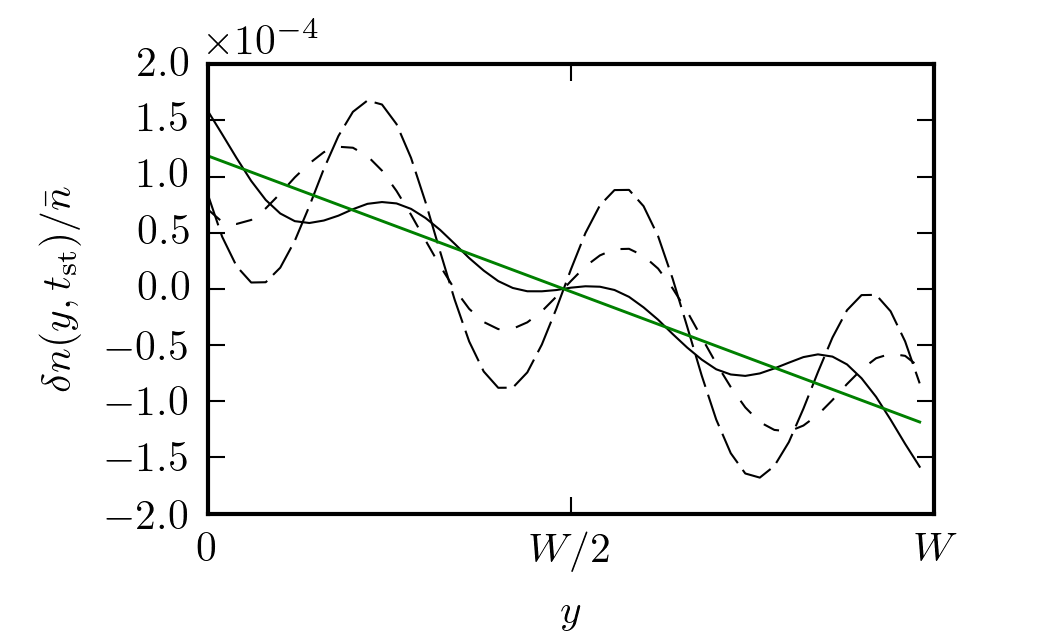}
\caption{\label{fig:plasmawaves} (Color online)
Time-evolution of the carrier density.
(a) The color plot displays the value of the density fluctuation $\delta n(y,t)$ (in units of the equilibrium density $\overline{n}=10^{11} \, \rm{cm}^{-2}$) as a function of time $t$ (in units of the driving period $T$) and as a function of the coordinate along the transverse direction $\hat{\bm y}$ of the ribbon.
(b) The black lines show the space profile of the density fluctuation at several times $t = 5.5 \, T$ (short-dashed line), $5.6 \, T$ (long-dashed line), and $5.7 \, T$ (solid line).
These times are marked by vertical lines with corresponding dashing in (a).
The green solid line is the value of the time-average of the density fluctuation at each point of the graphene bar. }
\end{figure}

In the previous sections, we used the value of the carrier density fluctuation at the upper and lower edges of the graphene ribbon, obtained by solving Eqs.~(\ref{eq:dyn-vis}), to calculate the Hall potential via the local-capacitance approximation [see Eqs.~(\ref{eq:Vh-1}) and~(\ref{eq:lca})].
However, it is also interesting to look at the spatial profile of the density fluctuation along $\hat{\bm y}$.
Fig.~\ref{fig:plasmawaves} shows periodic oscillations $\delta \tilde{n}(y,t)$ in both the space and time coordinates, ($y$ and $t$, respectively,) on top of a constant linear density slope $\langle \delta n(y, t) \rangle$ which decreases from the bottom to the top edge of the ribbon.
The total density fluctuation is given by the superposition of a constant and an oscillating term, $\delta n (y, t) = \langle \delta n(y, t) \rangle + \delta \tilde{n}(y,t)$.
Oscillations take place at an angular frequency $\omega_{\rm P} = 4 \pi f$, corresponding to \emph{twice} the driving frequency, as for the Hall electric potential (see Fig.~\ref{fig:hallvoltage}). 
The density slope is responsible for the time-averaged  Hall field along $\hat{\bm y}$ and its steepness determines the Hall voltage.

In the following, we demonstrate that the periodic oscillations of the density can be identified as {\it plasma waves} propagating along the transverse direction $\hat{\bm y}$ of the ribbon.
Plasma waves are the collective modes of the two-dimensional (2D) electron liquid~\cite{giuliani_vignale,maier_book} hosted by the graphene ribbon in the presence of a back gate.
In the long-wavelength $q \to 0$ limit, the energy dispersion $\hbar\omega(q)$ of standard 2D plasmons is proportional to the square root of the wave vector $q$~\cite{giuliani_vignale,maier_book}.
On the contrary, plasma waves feature a linear (``acoustic'') dispersion~\cite{principi_solidstate_2011,stauber_newjourphys_2012,svintsov_jap_2012} because of the presence of the back gate.
Screening exerted by free charges in the back gate cuts off the long-range tail of the Coulomb interaction between electrons and thus reduces the energy of (``softens'') the collective modes.
Plasmons and plasma-waves in graphene are relatively long-lived, due to the reduced impurity scattering, and can be frequency-tuned over a large range (from the THz to the MIR) by varying the average carrier density $\bar{n}$ by means of the back-gate.~\cite{grigorenko_naturephoton_2012,reserbat_acsphoton_2021}
Our analysis reveals that the combined action of the electric and magnetic fields at the hot spot is to {\it launch} plasma waves along the transverse direction $\hat{\bm y}$ of the ribbon.

\begin{figure}
{(a)}\includegraphics[width=0.98\columnwidth]{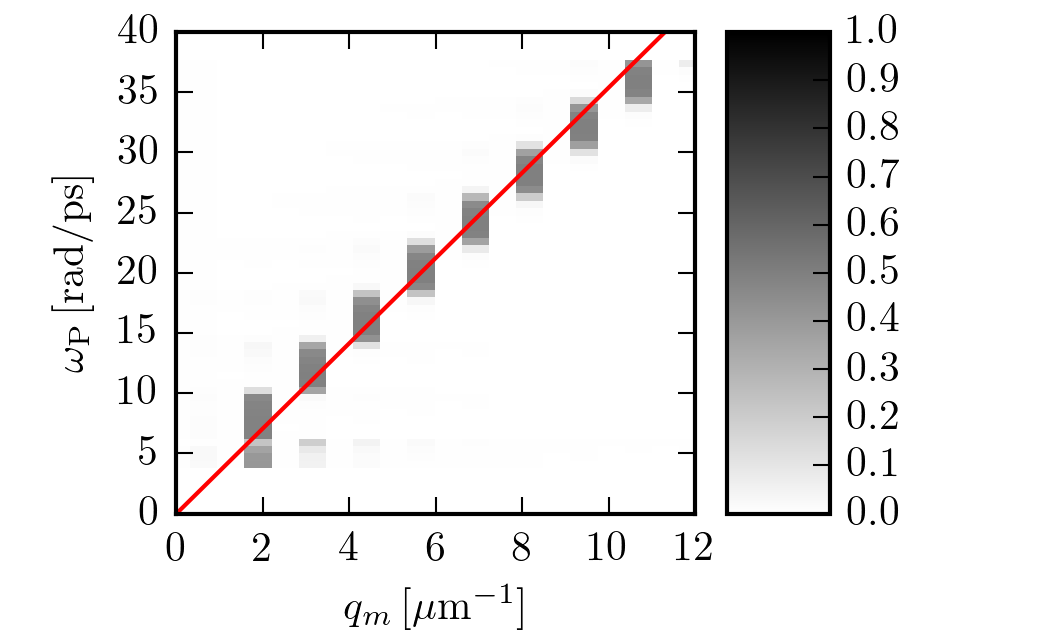} \\
{(b)}\includegraphics[width=0.98\columnwidth]{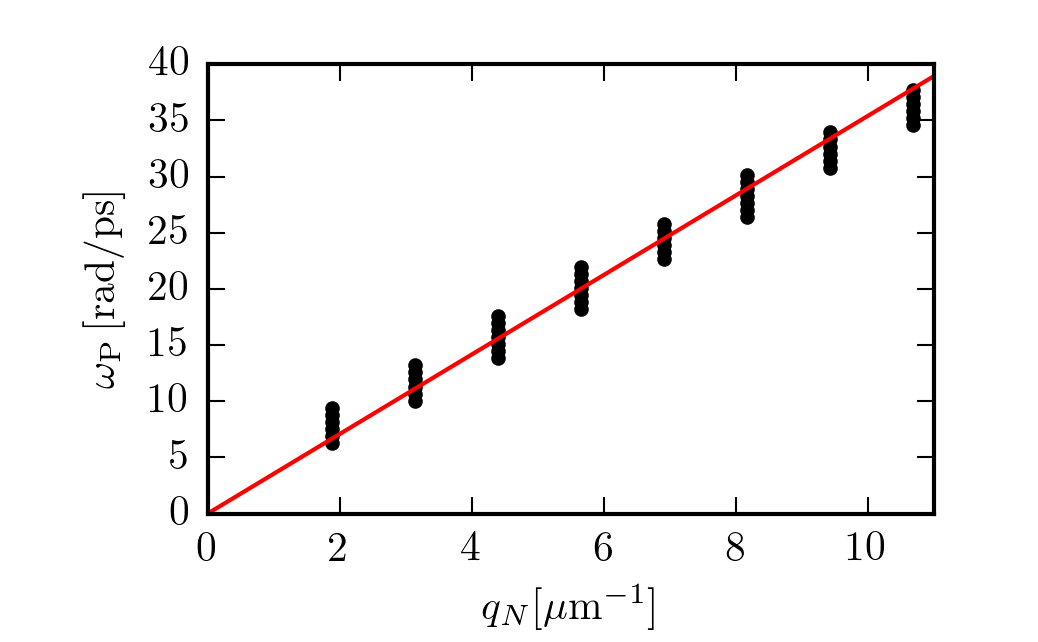}
\caption{\label{fig:dispersion}(Color online) 
Dispersion of the density fluctuations.
(a) Power spectrum $S(q_{m})$ of the density fluctuations along $\hat{\bm y}$, calculated from Eq.~(\ref{eq:spectrum}) as a function of $q_{m}$ and of the driving frequency $f = \omega_{\rm P} / (4 \pi)$.
(b) Wave vector $q_{N}= (\pi N)/ W$, defined as the number of nodes $N$ of the profile of the density fluctuations along $\hat{\bm y}$.
In both panels, the red solid line corresponds to the plasma-wave dispersion in Eq.~(\ref{eq:acoustic}).
For our choices of parameters (see first paragraph of Sec.~\ref{sec:numresults}) the plasma-wave speed is $s = 3.54  \, {\rm nm} / {\rm fs}$. }
\end{figure}

To substantiate our statement on the nature of the density oscillations, we calculate the Fourier transform $\delta \hat{n}(q_{m})$ of $\delta \tilde{n}(y,t)$ with respect to $y$,
\begin{equation}\label{eq:dft}
\delta \hat{n}(q_{m}) = \frac{1}{W} \int_{0}^{W} dy \, \delta \tilde{n}(y, t) e^{-i q_{m} y}~,
\end{equation}
at a large time $t = p T$ with $p \sim 100$, for a discrete set of wave vectors $q_{m} = m \pi / W$.
(We remind the reader that $T$ is the period of the external driving.)
The Fourier spectrum 
\begin{equation} \label{eq:spectrum}
S(q_{m})= \frac{|\delta \hat{n}(q_{m})|^{2}}{\sum_{m}|\delta \hat{n}(q_{m})|^{2}}
\end{equation}
is represented in Fig.~\ref{fig:dispersion}(a) as a density plot.
Each horizontal slice of the plot shows the Fourier spectrum, on the discrete set of wave vectors $q_{m}$, at a fixed driving frequency $f$.
The spectrum is represented as a piecewise constant function over segments of width $\pi/W$, centered at each wave vector $q_{m}$.
Only wave vectors corresponding to an odd $m$ have a sizable spectral weight, which is due to fluctuations having a node at $y = W/2$ but not at the edges $y = 0$, $L$ [see Fig.~\ref{fig:plasmawaves}(a)].  
The solid line in the plot shows the expected acoustic dispersion of plasma waves in the graphene ribbon, i.e.
\begin{equation}\label{eq:acoustic}
\omega_{\rm P}= s q, \quad \mbox{with} ~ s = \sqrt{e^{2} \bar{n}/(C m_{\rm c})}~.
\end{equation}
For each discrete wave vector $q_{m}$, the maximum of the Fourier spectrum is obtained for a driving frequency $f = s q_{m} / (4 \pi)$, corresponding to the plasma wave with that wave vector.
We reiterate that the extra factor of $2$ in the denominator is due to the fact that the electron density oscillates with \emph{twice} the frequency $f$ of the driving field.

Fig.~\ref{fig:dispersion}(b) reconstructs the dispersion of the density oscillations by relating the number $N$ of nodes in the profile of $\delta \tilde{n}(y, t)$ to the driving frequency.
The number of nodes is expressed in terms of the wave vector $q_{N} = (\pi N) / W$.
As in panel (a), we obtain that the dispersion coincides with the plasma waves of the graphene ribbon.

\section{Discussion}
\label{sec:discussion}

The main result of this work is that a graphene ribbon placed in a magnetic hot spot of a nanoassembly acts as a rectifier, converting the oscillating driving fields into a DC signal. 
The rectification is due to the response of the electron gas hosted in the graphene ribbon, which oscillates at the frequency $f$ of the driving fields $E_{\rm L}(t)$, $B_{\rm R}(t)$, \emph{and its harmonics}, and in particular at the $0{\rm th}$-order harmonic (which represents a constant density displacement).

To understand this rectification mechanism, let us illustrate in more detail why the harmonics of $f$ appear in the spectrum of the observables.
The current $J_{x}(y,t)$ is proportional to $E_{\rm L}(t)$ in Eq.~(\ref{eq:jx}), and thus its frequency spectrum contains $f$.
Because of the product $J_{x}(y,t)B_{\rm R}(t)$ in Eqs.~(\ref{eq:jy}), which represents the Lorentz force, the frequency spectrum of $J_{y}(y,t)$ contains both frequencies $f \pm f = 0,2 f$.
Then, with a similar argument applied to Eq.~(\ref{eq:jx}), the spectrum of $J_{x}(y,t)$ contains the frequency $f + 2 f = 3 f$ as well.
Iterating these considerations, it follows that the frequency spectrum of the currents contains all the harmonics of $f$.
The same happens to the Hall electric field and the Hall electric potential. 
In particular, the Hall voltage is just the $0{\rm th}$-order harmonic of the difference of the Hall electric potential between the upper and lower edges of the graphene ribbon. 

It is important to notice that, because of the hierarchical harmonics generation described above, the $0{\rm th}$-order harmonic includes contributions from arbitrarily large powers of the driving fields. 
In other words, the rectification mechanism is highly nonlinear.
In the derivation of Eq.~(\ref{eq:Vh-2}), by neglecting the Lorentz force in Eq.~(\ref{eq:jx}), we retain only the contribution of the second order nonlinearities to the $0{\rm th}$ order harmonic.

A similar hierarchy of harmonic generation is present also in the Navier-Stokes equations~(\ref{eq:dyn-vis}), where the velocity plays the role of the current. 
The numerical solution of the linearized hydrodynamic model does not limit the order of the nonlinear contributions and thus arbitrary harmonics can contribute to the time evolution of the variables. 
However, as Figs.~\ref{fig:hallvoltage},~\ref{fig:velocity}, and~\ref{fig:plasmawaves} illustrate, after an initial transient, the dynamics can be described in terms of $0 {\rm th}$ (Hall voltage), $1 {\rm st}$ [$v_{x}(t)$], and $2 {\rm nd}$ [$v_{y}$(t) and $\delta \tilde{n}(y,t)$] harmonics only.
For this reason, as Fig.~\ref{fig:largefreqs} shows, the agreement between the analytical and numerical calculation of the Hall voltage is excellent in a large frequency range.

It is useful to contrast the rectification mechanism described above with the well-known Dyakonov-Shur (DS) photodetection scheme, which also involves plasma waves in a driven 2D electron liquid.~\cite{dyakonov_ieee_1996,tomadin_prb_2013}
The DS scheme is based on hydrodynamic nonlinearities, which are intrinsic to the hydrodynamic equations of motions describing electron liquids, i.e.~the product of density and velocity in the continuity equation and the convective derivative in the Navier-Stokes equation.
On the contrary, the mechanism discussed here stems from \emph{linearized} hydrodynamic equations (see Sec.~\ref{sec:linearhydro}) and is thus fundamentally different from the DS scheme.

We also point out that, in our setup, the Hall voltage arises because of purely {\it classical} interactions, in contrast to  Hall-like nonlinear quantum effects in 2D and 3D materials \cite{moore_prl_2010,sodemann_prl_2015}, arising from a Berry curvature dip.

\section{Summary}
\label{sec:summary}

In conclusion, in this work we have shown that a graphene ribbon placed in the magnetic hot spot of a nanoassembly can be used as a magnetic field gauge.
The graphene's carriers are subject to the electric field driving the nanoassembly and to the magnetic field produced by it.
The carrier's response to the driving fields generates a constant potential difference (the Hall voltage) between two opposite edges of the ribbon.
The magnetic field, oscillating at optical frequencies, is thus rectified into a DC electrical signal which can be measured by a voltmeter connected to contacts at the ribbon's edges.
We have found a compact expression for the Hall voltage, Eq.~(\ref{eq:Vh-2}), by resorting to an approximate solution of the coupled equations for the current densities.
We have also numerically solved the linearized hydrodynamic equations for the two-dimensional electron liquid in the graphene ribbon, finding excellent agreement with the predictions based on Eq.~(\ref{eq:Vh-2}).
Finally, we have shown that, on top of the carrier's density slope generating the Hall voltage, standing plasma waves form across the ribbon.

\vphantom{A}
\vphantom{A}

\begin{acknowledgments}
This work was supported by the European Union's Horizon 2020 research and innovation programme under grant agreement no.~881603 - GrapheneCore3.
\end{acknowledgments}


\begin{thebibliography}{77}
\bibitem{footnote_nonlinear}
We have checked, by solving the nonlinear continuity and Navier-Stokes equations, that the nonlinearity in the divergence of the continuity equation and the convective nonlinearity in the Navier-Stokes equations do not significantly affect our results in the range of parameters explored here.
\end{thebibliography}
\end{document}